\newcommand{\PaperTitle}{Network and Systems Performance Characterization of MCP-Enabled LLM Agents}
\newcommand{\PaperNumber}{437}

\documentclass[10pt,sigconf,letterpaper,nonacm]{acmart}
\usepackage{booktabs}
\usepackage{graphicx}
\usepackage{amsmath}
\usepackage{balance}
\usepackage{siunitx}
\usepackage{enumitem}
\usepackage{tikz}
\usepackage{amsmath}
\usepackage{stfloats}
\usepackage{xcolor}

\usetikzlibrary{positioning, fit, shapes.geometric}

\begin{document}

\title{\PaperTitle}

\author{Zihao Ding}
\authornote{\textcolor{blue!55!black}{%
This manuscript reports experiments completed on May 15, 2025. Because the Model Context Protocol ecosystem evolves rapidly, the results reflect the state of the ecosystem at that time. We release this version on arXiv to provide a time\-stamped record of our instrumentation and measurements. Experiment artifacts are available from the corresponding author upon reasonable request (zd75@rutgers.edu).}}
\affiliation{\institution{Rutgers University}
\city{Piscataway}
\state{NJ}
\country{USA}
}
\email{zd75@rutgers.edu}

\author{Mufeng Zhu}
\affiliation{\institution{Rutgers University}
\city{Piscataway}
\state{NJ}
\country{USA}
}
\email{mz526@rutgers.edu}

\author{Yao Liu}
\affiliation{\institution{Rutgers University}
\city{Piscataway}
\state{NJ}
\country{USA}
}
\email{yao.liu@rutgers.edu}

\begin{abstract}
Model Context Protocol (MCP) has recently gained increased attention within the AI community for providing a standardized way for large language models (LLMs) to interact with external tools and services, significantly enhancing their capabilities. 
However, the inclusion of extensive contextual information, including system prompts, MCP tool definitions, and context histories, in MCP-enabled LLM interactions, dramatically inflates token usage. 
Given that LLM providers charge based on tokens, these expanded contexts can quickly escalate monetary costs and increase the computational load on LLM services. This paper presents a comprehensive measurement-based analysis of MCP-enabled interactions with LLMs, revealing trade-offs between capability, performance, and cost. 
We explore how different LLM models and MCP configurations impact key performance metrics such as token efficiency, monetary cost, task completion times, and task success rates, and suggest potential optimizations, including enabling parallel tool calls and implementing robust task abort mechanisms. 
These findings provide useful insights for developing more efficient, robust, and cost-effective MCP-enabled workflows.
\end{abstract}

\maketitle

\section{Introduction}
\label{sec:intro}

Large language models (LLMs) have surged in popularity in recent years, enabling everyday users to interact with them via chat-based interfaces. More recently, AI agents powered by LLMs also mark a significant advancement, capable of autonomous completion of complex tasks~\cite{luo2025large}. 
To facilitate interoperability among LLMs and external tools or APIs, the Model Context Protocol (MCP) was proposed as a standard by Anthropic~\cite{AnthropicMCPDocs,MCPIntroduction}. Since its introduction in November 2024, it has attracted considerable interest within the AI community. 
To date, website \texttt{mcp.so} indexes more than 13,000 MCP servers, allowing LLMs to access a diverse range of tools such as Github, Redis, Slack, Blender~\cite{AhujasidBlenderMCP}, and Unity~\cite{BarnettUnityMCP}.

While MCP offers a promising step towards standardizing agent-tool interaction, its workflow introduces important performance considerations.
As LLM APIs are largely stateless services, MCP-enabled interactions rely on rich contextual information, including detailed system prompts, conversation history, and results from previous tool calls, being included in each LLM prompt. 
This approach, while ensuring LLM has the necessary context for planning and action, can result in significant token consumption.
This, in turn, increases monetary costs to users and places heavier computational load on the LLM service providers for processing these input prompts.
Furthermore, the increased computational load can increase the task completion time, a key performance metric. 

In this paper, we investigate the performance implications of MCP-enabled LLM agent interactions. 
We instrumented Cline, an open-source MCP Host, to conduct controlled experiments: executing a set of MCP-enabled tasks with a diverse range of LLMs, repeated multiple times to obtain more stable estimates of their performance distribution. 
The metrics collected from these experiments include token counts, monetary costs, task completion times, and task success rates. 
To establish a comparative baseline for token usage patterns, we also analyzed a 90-day LLM general usage trace from an LLM replay service, OpenRouter. 
Overall, this paper makes the following contributions:
\begin{itemize}[leftmargin=*, topsep=1pt, partopsep=0pt, itemsep=0pt, parsep=1pt]
\item We empirically quantify the exceptionally low completion-to-prompt (C/P) token ratios in MCP-enabled LLM interactions. This is a direct consequence of the substantial contextual information, including system prompts and interaction history, that must be sent with each input prompt, underscoring the prompt-heavy nature of MCP tool calls.
\item We describe a multi-dimensional performance characterization of LLMs in MCP-enabled workflows. Using radar charts, we provide a comparative analysis of LLMs in terms of token usage, total monetary token cost, task success rate, and task completion time. 
\item We identify inefficiencies in multi-tool usage scenarios where current LLMs typically return only a single tool call per API request. We highlight potentials for reducing cumulative prompt token usage and task completion time by enabling LLMs to output multiple tool calls for parallel processing by MCP clients. 
\item We document cases where LLMs become stuck during complex tasks, which can result in the generation of excessive completion tokens and lead to high monetary costs, thereby necessitating reliable task abort mechanisms. 
\end{itemize}

\section{Background}
\label{sec:background}

\subsection{Model Context Protocol (MCP)}
\label{subsec:mcp_overview}

\begin{figure}
    \centering
    \includegraphics[width=0.46\textwidth]{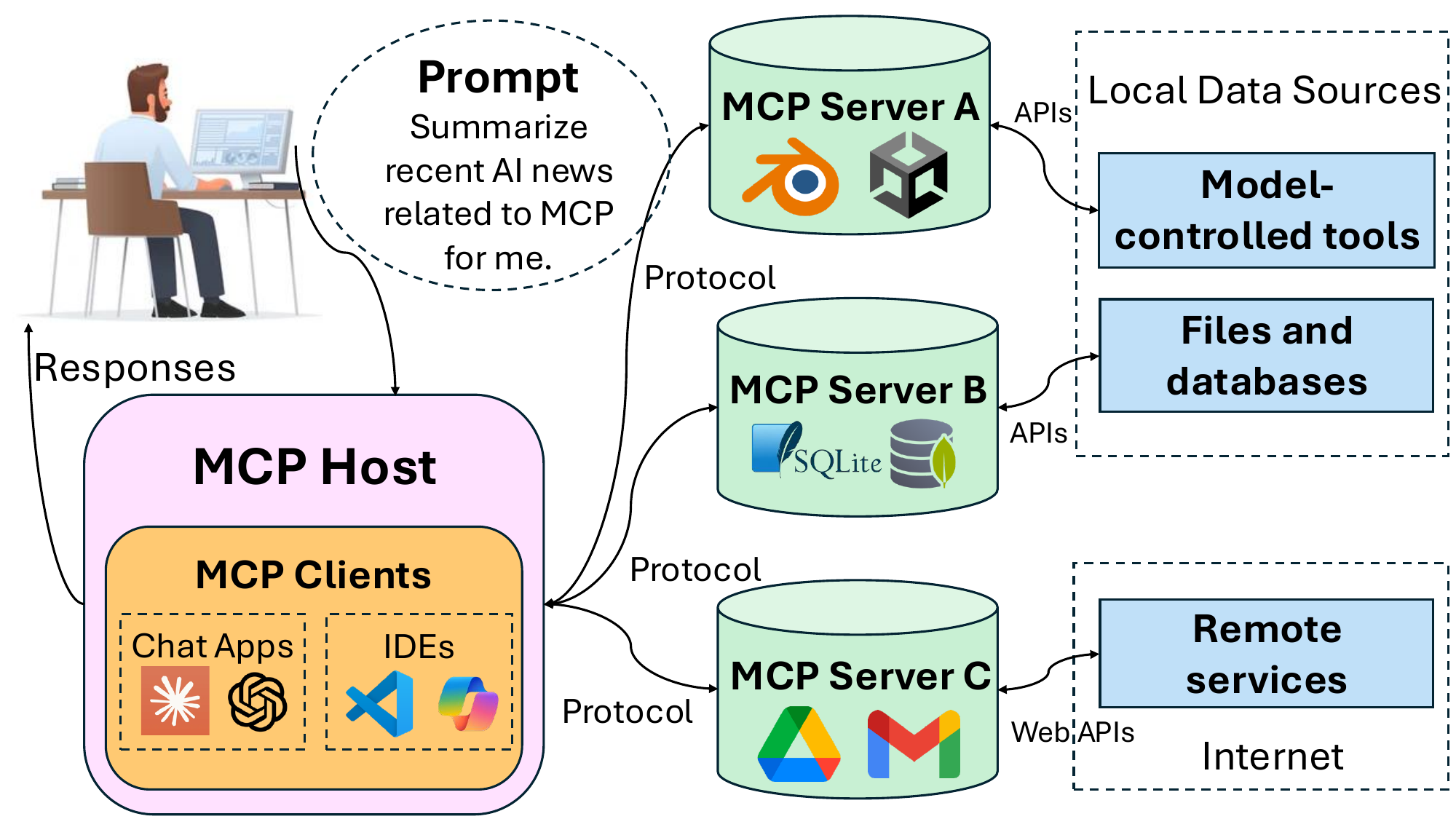}
    \vspace{-1em}
    \caption{MCP-enabled workflow}
    \vspace{-2em}
    \label{fig:mcp_architecture}
\end{figure}
Figure~\ref{fig:mcp_architecture} illustrates the primary parties involved in the execution flow of MCP-enabled LLM agents. These include the user initiating the task, the LLM providing reasoning capabilities, the underlying tool performing the action, and the core MCP components: the Host, Client, and Server.

\noindent\textbf{MCP Host} is the application that the user interacts with (e.g., an AI-enabled IDE~\cite{cline,cursor} or a desktop application like Claude Desktop~\cite{claudedesktop}). The Host receives user prompts, constructs the full context (including conversation history, system instructions, and information about available MCP tools/resources), communicates with the LLM, interprets the LLM's response (including potential requests to use a tool), and invokes the appropriate MCP Client to initiate tool execution.

\noindent\textbf{MCP Client} receives invocation requests from the Host when a specific MCP tool needs to be used. The Client handles the communication with the corresponding MCP Server according to the MCP specification (typically using JSON-RPC over a defined transport, e.g., with \texttt{stdio} or server-sent events (SSE) for server-to-client communication~\cite{mcptransports}. The Client translates the Host's request into MCP messages for the Server and relays the Server's response back to the Host.

\noindent\textbf{MCP Server} exposes specific capabilities (Tools, Resources, Prompts) via the MCP protocol. It is a RPC server that listens for connections from MCP Clients. Upon receiving a valid request (e.g., a tool call request), the Server invokes the underlying tool's API on behalf of the Client. The Server then sends the result back to the Client. 

\subsection{MCP Interaction Model and Context Management}
\label{subsec:mcp_interaction}

\begin{figure}[!t]
\centering
\begin{tikzpicture}[
    node distance=5mm and 3mm,
    mainbox/.style={
        draw,
        rectangle,
        thick,
        text width=0.46\textwidth,
        align=left,
        inner sep=3pt,
        font=\small
    },
    title/.style={
        font=\bfseries\small,
        anchor=south,
        inner sep=1pt
    },
    subbox/.style={
        draw,
        rectangle,
        thin,
        text width=0.42\textwidth,
        align=left,
        inner sep=2pt,
        fill=gray!10,
        font=\small
    }
    ]
    
    % 1. System Prompt Section
    \node (sys_prompt_title) [title] {System Prompt $(\mathcal{S})$};
    \node (sys_prompt_content) [mainbox, below=0.5mm of sys_prompt_title] {
        \begin{itemize}[leftmargin=*, topsep=1pt, partopsep=0pt, itemsep=0pt, parsep=1pt, label=\textbullet]
            \item Persona definition of the assistant
            \item Tool use instructions, examples, guidelines
            \item \textbf{Introduction to MCP framework, connected MCP servers, detailed MCP manuals describing supported tools/resources}
            \item Defines the two operational modes: ACT and PLAN
            \item AI's capabilities and how tools can be combined
            \item Rules AI must follow
            \item Details about operating environment (OS, Shell, Paths)
            \item Instructions on how AI can solve the problem
        \end{itemize}
    };
    
    % 2. Interaction History Section
    \node (hist_title) [title, below=2mm of sys_prompt_content] {Interaction History $(\mathcal{H})$};
    \node (hist_content_box) [mainbox, minimum height=2cm, below=0.5mm of hist_title] {};
    
    \node (dots) [font=\footnotesize, below=1mm of hist_title, anchor=north east, xshift=-5pt] { 
        $\dots$ (repeat for turns $1..N-1$)
        };

    \node (turn_i_box) [subbox, below=1mm of dots, anchor=north west, xshift=-51pt] {
        \textbf{Turn $i$:}
        \begin{itemize}[leftmargin=*, topsep=0pt, partopsep=0pt, itemsep=0pt, parsep=0pt, label=\textbullet] % Minimal spacing
            \item \{\ User Prompt | Tool Call Result \} ($p_i$)
            \item LLM Output ($o_i$)
            % \item Tool Result ($t_i$)
            % \item LLM Task Outcome ($o_i$)
        \end{itemize}
    };
    
    % 3. Current Turn N Information Section
    \node (current_title) [title, below=2mm of hist_content_box] {Current Turn $N$ Information $(\mathcal{C})$};
    \node (current_content) [mainbox, below=0.5mm of current_title] {
        \begin{itemize}[leftmargin=*, topsep=1pt, partopsep=0pt, itemsep=0pt, parsep=1pt, label=\textbullet]
            \item Current \{\ User Prompt | Tool Call Result \} ($p_N$)
            % \item Current Environment Details ($e_N$)
        \end{itemize}
    };
    
     \node [draw, thick, solid, fit=(sys_prompt_title)(sys_prompt_content)(hist_title)(hist_content_box)(current_title)(current_content)] {};
    
\end{tikzpicture}
\vspace{-2em}
\caption{Structure of the API requests sent from Cline (the MCP Host) to the LLM}
\vspace{-2em}
\label{fig:cline_context}
\end{figure}
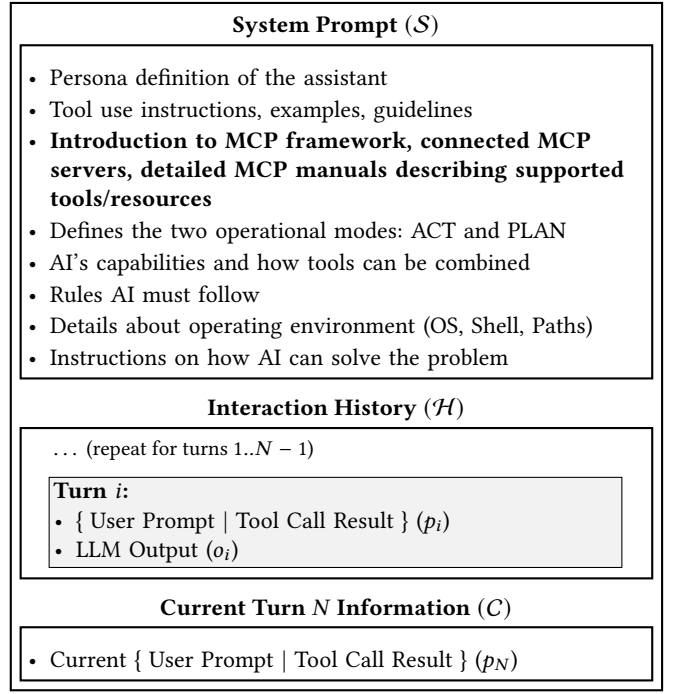

Taking the MCP Host ``Cline''~\cite{cline} as an example, Figure \ref{fig:cline_context} shows the full prompt constructed and sent by the Host to the LLM.
All information in the figure is included in the context window, and is sent as part of the API request to the model server. The total length of the context window sent in the $N$-th LLM API request, $\mathcal{L}_N$, can be obtained as $\mathcal{L}_N = \mathcal{S}_N + \mathcal{H}_N + p_N$, where $\mathcal{H}_N = \sum_{i=1}^{N-1} \left( p_i + o_i \right)$ represents the length of conversation history from the previous $N-1$ turns.

The total length of messages (both API request and responses) transmitted during a N-turn conversation can be represented as $\mathcal{T}_N = \sum_{i=1}^{N} \left( \mathcal{S}_i + \mathcal{H}_i  + p_i \right) + o_N$.

\subsection{MCP-Enabled vs. Chat-Based LLM Use}
\label{subsec:contrast}

Maintaining conversational context by including interaction history in the prompt is standard practice for stateful LLM applications. Because the underlying LLM API calls are typically stateless between requests, the client application manages the state by sending the relevant history within the context window. This stateless nature of the API allows flexibility, such as  switching the underlying model or model server during a conversation if the client application is designed to support it~\cite{yu2025stateful,martin2024llmproxy}. State information, including the chat history, is sent with each API call within the context window, enabling the model to generate responses based on the current turn and past interactions.

In typical multi-turn conversational chatbot interactions (e.g., via interfaces like ChatGPT), the number of tokens generated in the LLM's response significantly exceeds the number of tokens in the user's prompt and the accompanying history sent by the client. This establishes a baseline expectation for conversational token usage. Table~\ref{tbl:chatbot-llm-stats} presents statistics compiled from several large-scale public datasets of such conversations~\cite{zheng2023lmsys,zhao2024wildchat}. As shown, the average number of response tokens is consistently multiple times larger than the average number of prompt tokens across these diverse datasets, with calculated average Response-to-Prompt token ratios ranging from approximately 1.5 to 6.3. This baseline contrasts sharply with the token usage patterns we observe in MCP-enabled agents, as detailed in Section~\ref{sec:results}.

\begin{table}[!t]
\small
  \centering
\caption{Selected Statistics for Conversational Chatbot Datasets (Data compiled from~\cite{zheng2023lmsys,zhao2024wildchat})}
\vspace{-1em}
  \label{tbl:chatbot-llm-stats}
\scalebox{0.88}{
  \begin{tabular}{l S[table-format=1.1] S[table-format=2.1] S[table-format=3.1]} 
    \toprule
    Dataset & {Avg. \# Turns} & {Avg. \# Tokens} & {Avg. \# Tokens} \\ 
     & {per Sample} & {per Prompt} & {per Completion} \\
    \midrule
    Anthropic HH~\cite{bai2022training}    & 2.3 & 18.9 &  78.9 \\ % 4.2
    OpenAssistant~\cite{kopf2023openassistant}   & 2.3 & 33.4 & 211.8 \\  % 6.3
    Chatbot Arena~\cite{zheng2023judging}   & 1.2 & 52.3 & 189.5 \\ % 3.62
    LMSYS-Chat-1M~\cite{zheng2023lmsys}   & 2.0 & 69.5 & 214.5 \\ % 3.1
    WildChat~\cite{zhao2024wildchat} & 2.5 & 295.6 & 441.3 \\ % 1.5
    \bottomrule
  \end{tabular}
  }
  \vspace{-2em}
\end{table}

\section{Methodology}
\label{sec:methodology}

For our measurement study, we report and compare data collected from two sources: (1) a 90-day usage trace from OpenRouter~\cite{OpenRouterAI} – a commercial relay service that forwards end-user requests to various Large Language Model (LLM) backends and provides detailed usage metrics; and (2) data from a controlled benchmark conducted using Cline, an MCP Host that we instrumented by modifying its source code to enable comprehensive logging of API interactions.

\subsection{Key Performance Metrics}
\label{sec:metrics}

In this study, we collect and derive the following comprehensive set of performance metrics to evaluate the performance and effectiveness of MCP-enabled LLM interactions:

\noindent\textbf{\# of Prompt Tokens} quantifies the length of input to the LLM. 
The prompt tokens include not only the direct user input but also system prompt and the preceding interaction history, as illustrated in Figure \ref{fig:cline_context}. 
The average \# of tokens per word is model-dependent. For example, OpenAI has reported an approximate average of 1.33 tokens per word~\cite{OpenAITokensHelp}. 

\noindent\textbf{\# of Completion Tokens} measures the length of textual output generated and returned by the LLM in response to an API request.

\noindent\textbf{Token Efficiency} is the total number of tokens consumed by the LLM to complete a task, reflecting resource utilization.

\noindent\textbf{Cost per Task} is the monetary cost charged by LLM APIs for completing a task. This cost is charged for both the input (prompt tokens) and output (completion tokens), often priced per one million tokens. 
While token costs across different LLMs and service providers vary significantly, completion tokens are typically priced substantially higher -- 4x to 8x times -- than prompt tokens. 

\noindent\textbf{Task Success Rate} measures the proportion of tasks successfully completed by the LLM in the MCP-enabled workflow.

\noindent\textbf{Task Completion Time} measures the total time taken for the LLM to complete a task. A task may include multiple turns/steps, each including an LLM API request/response.

\subsection{OpenRouter 90-Day Usage Trace}
\label{sec:openrouter_trace}

To quantify how MCP shifts token-flow style in live deployments, we obtained a full 90-day export of token usage logs from OpenRouter. Our analysis focused on the \texttt{deepseek}, \texttt{qwen}, \texttt{llama}, \texttt{openai chatgpt}, \texttt{google gemini}, and \texttt{anthropic claude} model families, which were identified as the most popular models on the OpenRouter platform during the data collection period based on both paid and free usage metrics. 
The data is from Feb 5, 2025 to May 5, 2025.

For each model, the trace reports the daily aggregated total \# of prompt tokens, total \# of completion tokens, and the \# of requests. This allows us to derive the average \# of prompt and completion tokens per request for each model. 
However, we also note that the trace
does not include per-user or per-request information. As a result, our analysis of this trace is limited to the ``general usage patterns'' observed on the OpenRouter platform. 
We use this general usage pattern as a baseline for comparison with the token usage characteristics observed in our controlled MCP-enabled experiments.

\subsection{Controlled Benchmarking with Instrumented Cline}
\label{sec:cline_setup}

We use Cline~\cite{cline}, an open-source tool designed for software development, to serve as the AI agent and MCP Host. 
To facilitate experiments across a diverse range of LLMs, we configure Cline v3.14.0 to interact with different LLMs, including \texttt{deepseek}, \texttt{qwen}, \texttt{llama}, \texttt{openai chatgpt}, \texttt{google gemini}, and \texttt{anthropic claude}, via OpenRouter.
We created a set of MCP-driven tasks of varying complexities. 
We acknowledge that prompt engineering plays a crucial role in influencing LLM performance. To ensure a fair comparison, we carefully designed and refined our prompts through extensive testing.

To record detailed performance data, we modified the Cline source code to implement comprehensive logging of API interactions with the LLM servers, including the \# of prompt tokens, the \# of completion tokens, \# of API request/responses, task completion time, and token cost. Table \ref{tab:model_performance_summary} summarizes the key performance metrics derived from a total of 820 completed tasks.

\begin{table}[!t]
    \centering
    \caption{Abbreviated Model Names for Plots and Tables}
\vspace{-1em}
    \scalebox{0.9}{
    \begin{tabular}{ll}
        \toprule
        \textbf{Full Model Name} & \textbf{Abbreviation} \\
        \midrule
        Anthropic: Claude 3.7 Sonnet & C3.7S \\
        DeepSeek: DeepSeek V3 0324 & DSv3-0324 \\
        DeepSeek: R1 & DS-R1 \\
        DeepSeek: R1 Distill Llama 70B & DS-R1-Distill-70B \\
        Google: Gemini 2.0 Flash & G2.0F \\
        Google: Gemini 2.0 Flash Lite & G2.0F-Lite \\
        Google: Gemini 2.5 Flash Preview & G2.5F-Prev \\
        Meta: Llama 4 Maverick & Llama4-Mav \\
        OpenAI: GPT-4o-mini & GPT-4o-mini \\
        Qwen: Qwen3 235B A22B & Qwen3-235B-A22B \\
        Qwen: Qwen3 30B A3B & Qwen3-30B-A3B \\
        \bottomrule
    \end{tabular}
    }
    \vspace{-1em}
\end{table}

\begin{figure}[!t]
    \centering
    \includegraphics[width=0.46\textwidth]{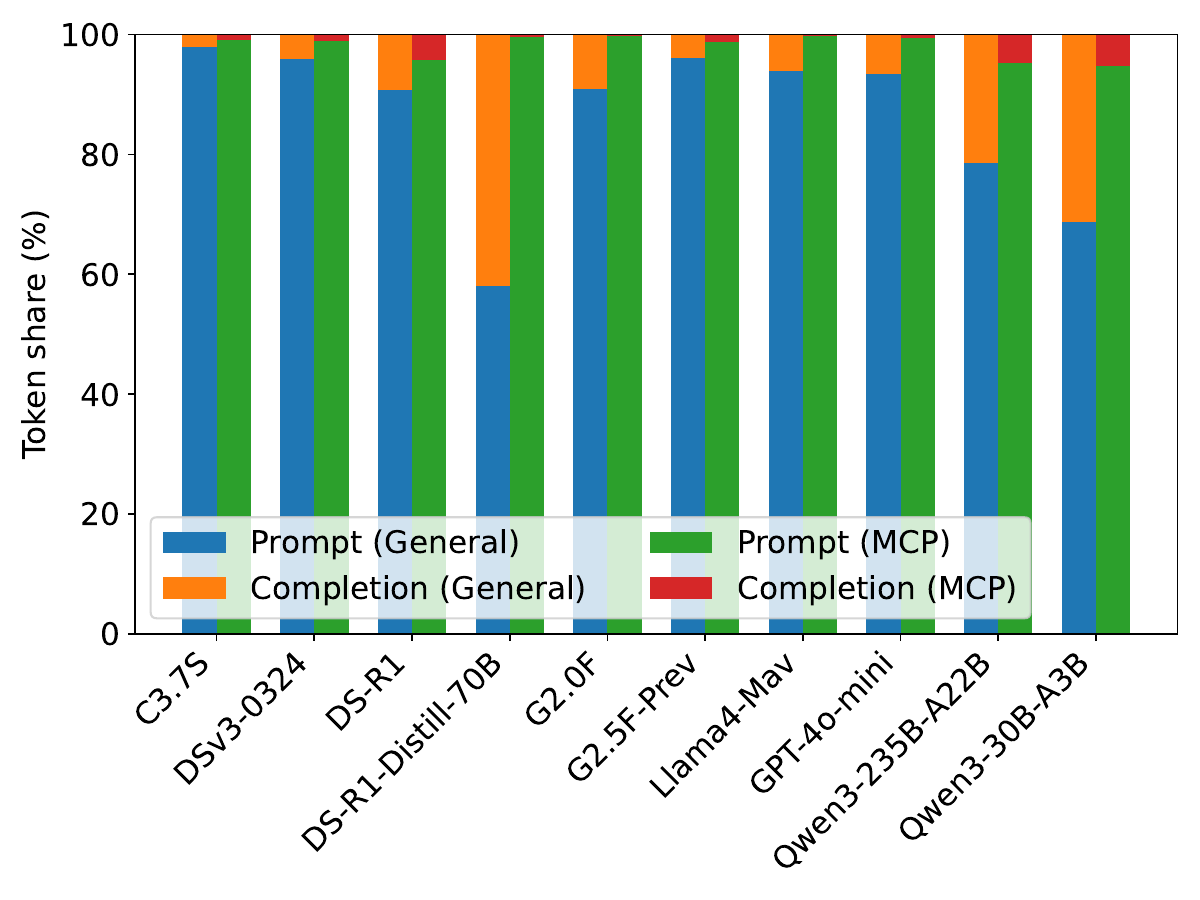}
    \vspace{-2em}
    \caption{Prompt Token vs. Completion Token.}
    \vspace{-1em}
    \label{fig:prompt_completion_ratio_comp} 
\end{figure}

\section{Token Usage Patterns: General vs. MCP-Enabled}

Figure \ref{fig:prompt_completion_ratio_comp} compares the \% of prompt tokens (input) versus completion tokens (output) across different LLMs.
For each LLM represented on the x-axis, the left bar reports the ``general usage pattern'' derived from the 90-day OpenRouter trace, while the right bar reports the ``MCP-enabled usage pattern'' obtained from our instrumented Cline benchmark.
Table \ref{tab:my_token_flow_comparison} provides  more detailed statistics for a selection of these models.

Our analysis of the 90-day OpenRouter trace reveals consistently low ``completion-to-prompt'' (C/P) ratios, defined as the \# of completion tokens divided by the \# of prompt tokens, across all examined models. The observed C/P ratios in this general usage data range between $0.021$ and $0.722$, with a median value of $0.111$. 
Models with chain-of-thought reasoning abilities such as DeepSeek R1, Anthropic Claude-3.7-sonnet:thinking, and Qwen3-30B-A3B, where the thinking chain processes are counted as part of the completion token, tend to have higher C/P ratios. 
On the other hand, even among these ``thinking'' models, those with extra large context window sizes (e.g., \texttt{gemini} models) can have low C/P ratios (e.g., $0.075$). 
This is likely because the large context windows can accommodate longer input prompts.

In contrast, our analysis of token usage data for MCP-enabed tasks reveals significantly different C/P ratios. 
Specifically, the C/P ratios for MCP interactions are 2x-30x lower than those observed in general OpenRouter traffic. This confirms that MCP host communications are inherently prompt-heavy, where the vast majority of tokens are consumed as input instructions rather than LLM's generated output.

The main reason for this prompt-heavy profile is the nature of MCP's task orchestration.
As shown in Figure \ref{fig:cline_context}, for each LLM API request in an MCP-enabled workflow, the host application need to transmit a detailed system prompt, the interaction history, and observations from previous tool calls. 
As shown in Table \ref{tab:my_token_flow_comparison}, even models with high C/P ratios under general usage show significant reductions when used in an MCP-enabled setting. 
This phenomenon underscores the LLM's role in MCP-enabled workflow, which is to process extensive contextual input to make tool call decisions rather than to generate lengthy free-form text. 

\begin{table*}[htbp]
\centering
\caption{Sample Comparison of Token Usage Metrics: General Usage vs. MCP-enabled Tasks}
\vspace{-1em}
\label{tab:my_token_flow_comparison}
\small
\begin{tabular}{@{}lrrrrrrrr@{}}
\toprule
{\centering \textbf{Model Name}} & \multicolumn{1}{p{1.7cm}}{\centering General \\ C/P Ratio (\%)} & \multicolumn{1}{p{1.7cm}}{\centering MCP \\ C/P Ratio (\%)} & \multicolumn{1}{p{1.6cm}}{\centering General Avg. Prompt Tok.} & \multicolumn{1}{p{1.6cm}}{\centering MCP Avg. Prompt Tok.} & \multicolumn{1}{p{1.6cm}}{\centering General Avg. Compl. Tok.} & \multicolumn{1}{p{1.6cm}}{\centering MCP Avg. Compl. Tok.} & \multicolumn{1}{p{1.6cm}}{\centering General Avg. Total Tok.} & \multicolumn{1}{p{1.6cm}}{\centering MCP Avg. Total Tok.} \\
\midrule
C3.7S & 2.1 & 0.99 & 28918 & 27268 & 620 & 271 & 29538.4 & 27538 \\
DSv3-0324 & 4.2 & 1.07 & 7618 & 16512 & 322 & 177 & 7940.6 & 16689 \\
DS-R1 & 10.2 & 4.46 & 7550 & 15698 & 771 & 700 & 8321.1 & 16398 \\ 
G1.5F-8B & 10.8 & 0.31 & 833 & 13826 & 90 & 43 & 922.8 & 13869 \\ 
G2.0F & 9.9 & 0.18 & 2517 & 122331 & 250 & 220 & 2767.4 & 122551 \\ 
\bottomrule
\end{tabular}
\end{table*}

\begin{table*}[!t]
\centering
\caption{Summary of Model Performance Metrics in MCP-enabled Task Workflows}
\vspace{-1em}
\label{tab:model_performance_summary}
\small 
\begin{tabular}{@{}lrrrrrrrrr@{}}
\toprule
\multicolumn{1}{p{2.0cm}}{\centering\textbf{Model Name}} &
\multicolumn{1}{p{0.6cm}}{\centering\textbf{Total Reqs}} &
\multicolumn{1}{p{1.5cm}}{\centering\textbf{Total Prompt Tokens}} &
\multicolumn{1}{p{1.2cm}}{\centering\textbf{Total Compl. Tokens}} &
\multicolumn{1}{p{1.1cm}}{\centering\textbf{Avg. Req Cost (\$)}} &
\multicolumn{1}{p{1.8cm}}{\centering\textbf{Avg. Prompt Tokens per Task}} &
\multicolumn{1}{p{1.7cm}}{\centering\textbf{Avg. Compl. Tokens per Task}} & 
\multicolumn{1}{p{1.6cm}}{\centering\textbf{Avg. Total Tokens Task}} & 
\multicolumn{1}{p{1.5cm}}{\centering\textbf{Avg. Compl. Time (ms)}} \\ \midrule

C3.7S & 332 & 9,052,845 & 89,903 & 0.0538 & 164,597 & 1,635 & 166,232 & 70,911 \\
DSv3-0324 & 743 & 12,268,230 & 131,540 & 0.0047 & 77,647 & 833 & 78,480 & 85,248 \\
DS-R1-Distill-70B & 593 & 28,594,873 & 124,161 & 0.0049 & 697,436 & 3,028 & 700,464 & 244,712 \\
DS-R1-Distill-70B & 383 & 6,012,465 & 268,024 & 0.0094 & 58,373 & 2,602 & 60,976 & 126,143 \\
G2.0F & 537 & 65,691,519 & 118,204 & 0.0245 & 875,887 & 1,576 & 877,463 & 54,053 \\
Llama4-Mav & 834 & 62,292,273 & 100,394 & 0.0128 & 707,867 & 1,141 & 709,008 & 76,042 \\
GPT-4o-mini & 397 & 6,182,649 & 37,558 & 0.0014 & 89,604 & 544 & 90,148 & 32,376 \\
Qwen3-235B-A22B & 245 & 3,409,962 & 170,776 & 0.0015 & 68,199 & 3,416 & 71,615 & 146,709 \\
Qwen3-30B-A3B & 232 & 3,269,091 & 178,489 & 0.0016 & 46,044 & 2,514 & 48,557 & 51,844 \\
\bottomrule
\end{tabular}
\vspace{-1em}
\end{table*}

\section{Performance Characterization of MCP Interactions}
\label{sec:results}
This section presents a comprehensive evaluation of LLM performance with MCP servers. We introduce a novel tool-use benchmark, evaluating nine LLMs across three distinct MCP servers (Time~\cite{MCPTimeServerModule}, Playwright~\cite{MicrosoftPlaywrightMCP}, and Google Maps~\cite{MCPGoogleMapsModule}). Our benchmark assesses four key performance metrics, including task success rate, token efficiency, time for task completion, and price per task, to provide a nuanced understanding of LLM capabilities in interacting with MCP tools. Furthermore, we categorize the benchmark into simple and complex tasks to analyze the impact of task complexity on LLM performance and to expose performance variations between ``best'' and ``worst'' performing models.

\subsection{Benchmarking Tool-Use Efficiency}

To evaluate LLM tool-use efficiency, we designed a benchmark utilizing three MCP servers:

\begin{itemize}[leftmargin=*, topsep=1pt, partopsep=0pt, itemsep=0pt, parsep=1pt]
    \item \textbf{Time MCP:}  This server provides straightforward time-related information. Its simplicity makes it ideal for assessing basic MCP interaction. %(as of Apr 23, 2025, server installed via python)
    \item \textbf{Playwright MCP:} A server for web browser automation, presenting complexities like dynamic content and website structure. It allows AI models to interact with and understand web pages for tasks such as testing and automation. %(v0.0.20, server installed via npx) 
    \item \textbf{Google Maps MCP:} This MCP server enables location-based services, offering a blend of structured data and the intricacies of handling location-specific requests. %(as of Apr 10, 2025, server installed via npx)
\end{itemize}

Across these MCP servers, we designed prompts and categorized them into simple and complex based on task complexity. 
This allows us to evaluate the limitations and strengths of different LLMs working with MCP servers.

\begin{figure}[!t] 
    \centering
    \includegraphics[width=0.45\textwidth]{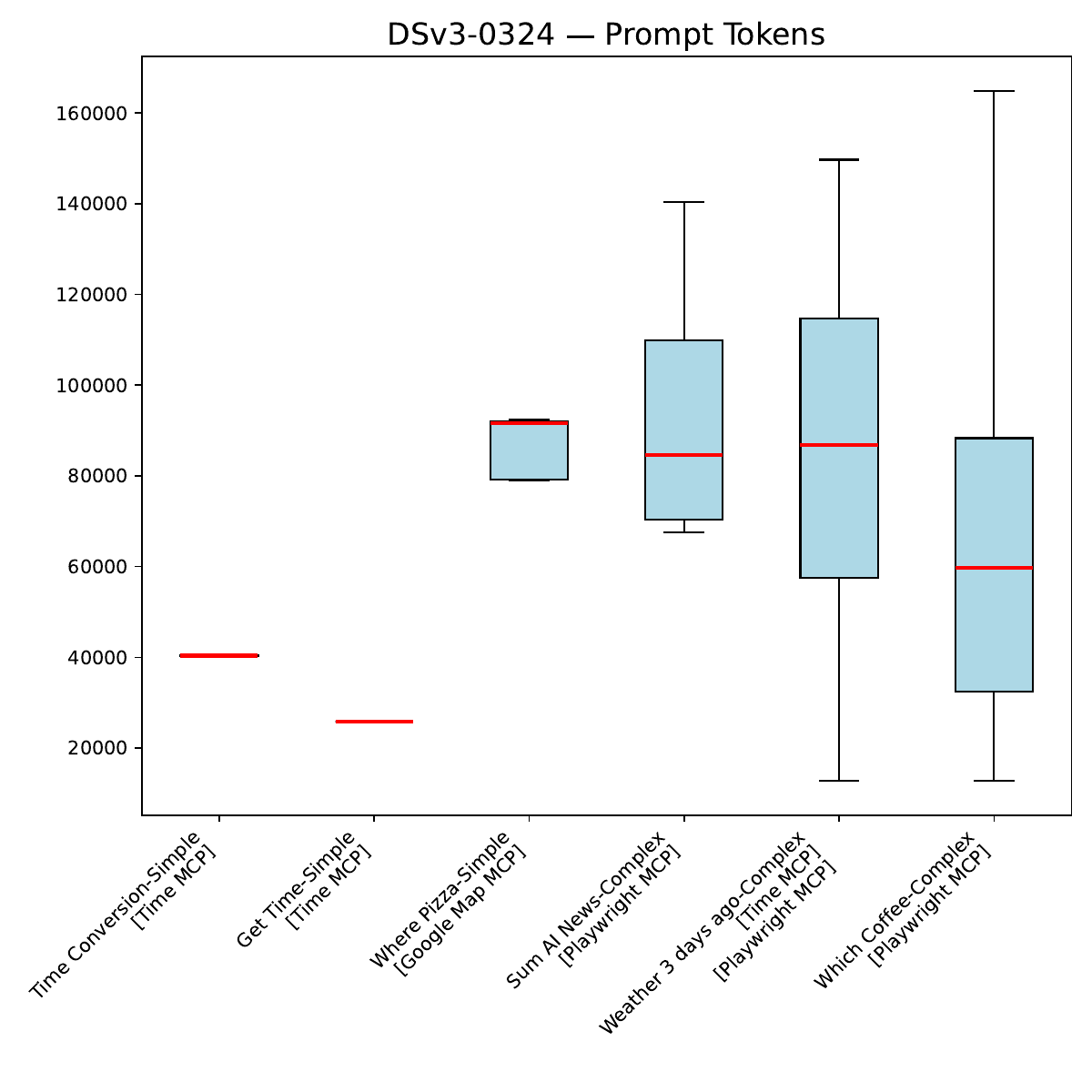}
    \vspace{-2.5em}
    \caption{Task Difficulty vs. Total Tokens}
    \label{fig:boxplot_task} 
    \vspace{-1.5em}
\end{figure}

\vspace{0.5em}
\noindent\textbf{Simple Tasks}
typically require minimal sequential steps and direct interaction with a single MCP tool,
with predictable and structured API responses. Examples include:

\begin{itemize}[leftmargin=*, topsep=1pt, partopsep=0pt, itemsep=0pt, parsep=1pt]
    \item Retrieving the current time using the Time MCP.
    \item Finding the nearest pizza shop using the Google Maps MCP at a specific location. 
    %(While involving location, the Google Maps API response is consistently formatted and predictable.)
\end{itemize}

\vspace{0.5em}
\noindent\textbf{Complex Tasks} generally
take multiple sequential steps and require the LLM to orchestrate interactions with several different MCP tools. These tasks may involve greater unpredictability and require  navigating more challenging environments, such as interacting with dynamic web page content and circumventing potential bot detection measures, demanding more advanced reasoning, planning, and error-handling capabilities. Examples include:

\begin{itemize}[leftmargin=*, topsep=1pt, partopsep=0pt, itemsep=0pt, parsep=1pt]
\item Searching for news, accessing a selected article's webpage, and then summarizing its content.%"Summarize AI news for me related to MCP on bing news. Open first link and summarize content" 
\item Automating a user login sequence or gathering data from a website using the Playwright MCP.
\end{itemize}

\vspace{0.1em}
The primary distinction between simple and complex tasks lies in the predictability and complexity of the MCP interactions. As result shown in Figure \ref{fig:boxplot_task}, the three ``simple'' MCP tasks (time conversion, get time, and where pizza) have virtually no inter-quartile range or whiskers on the box plot, their token counts are nearly identical on every run, whereas the three ``complex'' Playwright MCP tasks show dramatically larger boxes and whiskers, revealing high variations in total tokens. This contrast in distribution directly validates our taxonomy of simple versus complex tasks. 
Our initial observations revealed that web browsing tasks using the Playwright MCP presented significant challenges, due to LLMs' varying levels of prior knowledge, diverse and unpredictable web page content, tool use complexity, and websites' bot detection system.
These factors contribute to the increased difficulty of complex tasks, impacting the LLMs' ability to effectively utilize MCPs.
To quantitatively assess LLM performance, we measured the metrics described in Section \ref{sec:metrics}.
\vspace{-1em}

\subsection{Benchmark Results}
\begin{figure}[!t] 
    \centering
    \includegraphics[width=0.48\textwidth]{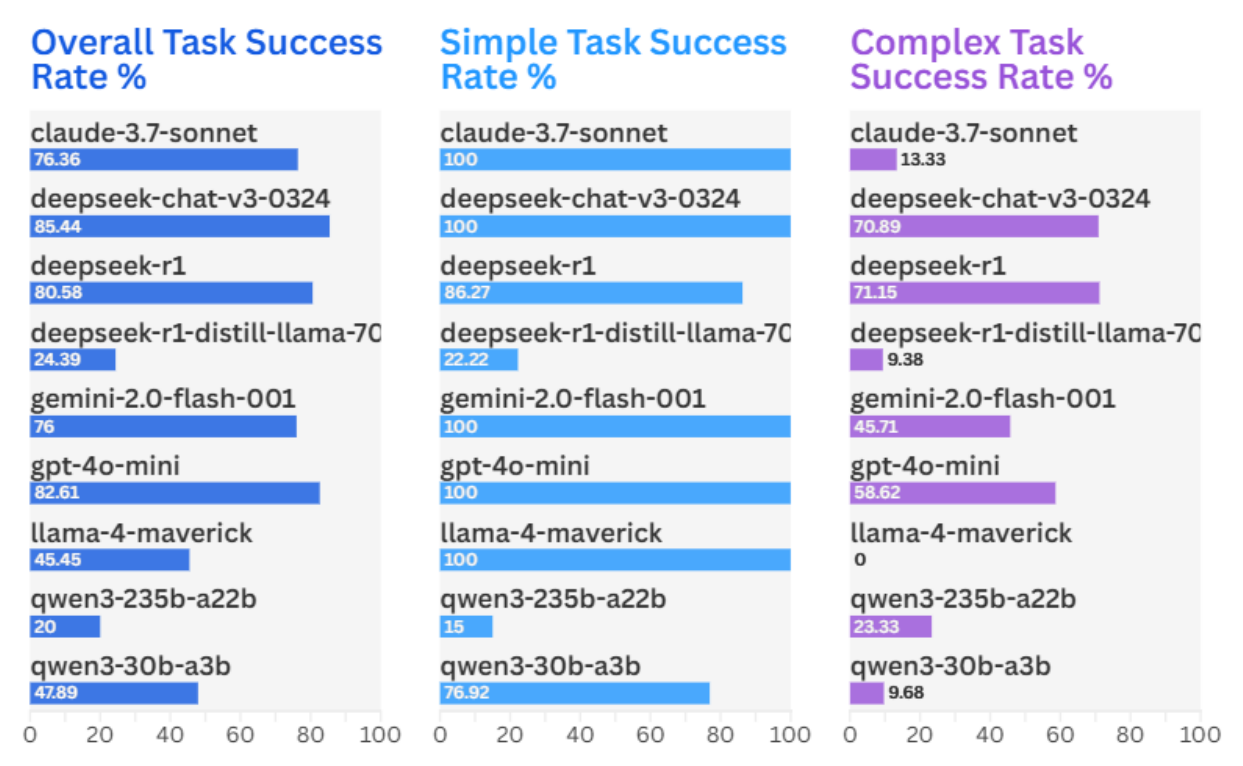}
    \vspace{-2em}
    \caption{Task Success Rate Comparisons.}
    \vspace{-1em}
    \label{fig:task_success_rate_by_models} 
\end{figure}

Figure \ref{fig:task_success_rate_by_models} reports the task success rates across different models. 
We find that most models exhibit strong performance on simple tasks, achieving high success rates. However, performance generally declines with complex tasks, highlighting the challenges of tool selection, interaction, and handling uncertainty. 

\begin{figure}[!t]
	\centering
	\includegraphics[width=0.32\textwidth]{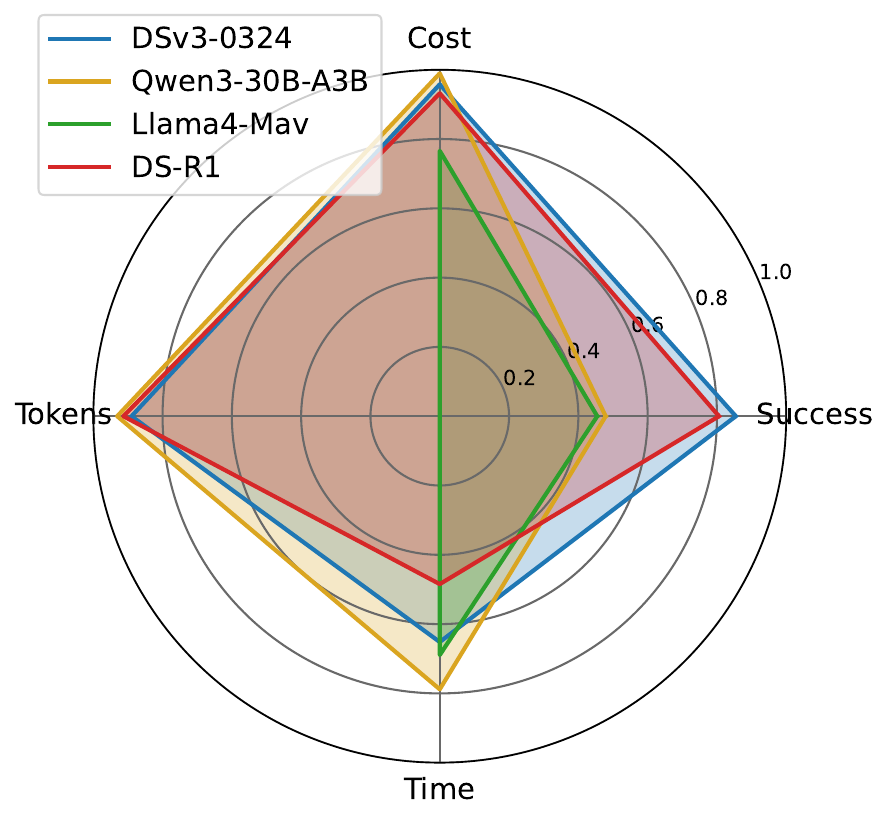}
	\vspace{-0.95em}
	\caption{Radar chart of MCP benchmark results, comparing token overhead, latency, and cost.}
	\label{fig:radar_chart}
	\vspace{-1.5em}
\end{figure}

To further analyze each model's capability with MCP-enabled workflows, we present our results as radar charts for clearer visualization in Figure~\ref{fig:radar_chart}, considering four key metrics: token efficiency, monetary cost, task completion times, and task success rates. 
(Extended visualizations appear in the Appendix: see the simple-task radar charts in Appendix Fig.~\ref{fig:simple_radar}, the complex-task radar charts in Appendix Fig.~\ref{fig:complex_radar}, and the average token consumption per task in Appendix Figs.~\ref{fig:total_tokens_per_task}.)
Notably, \texttt{gemini}-2.0 and 2.5 models performed well on complex tasks but with a higher token usage. Overall, \texttt{gpt}-4o-mini and \texttt{DeepSeek-V3} emerged as cost-effective models for MCP interaction in our testing.

\begin{figure}[!t]
	\centering
	\includegraphics[width=0.45\textwidth]{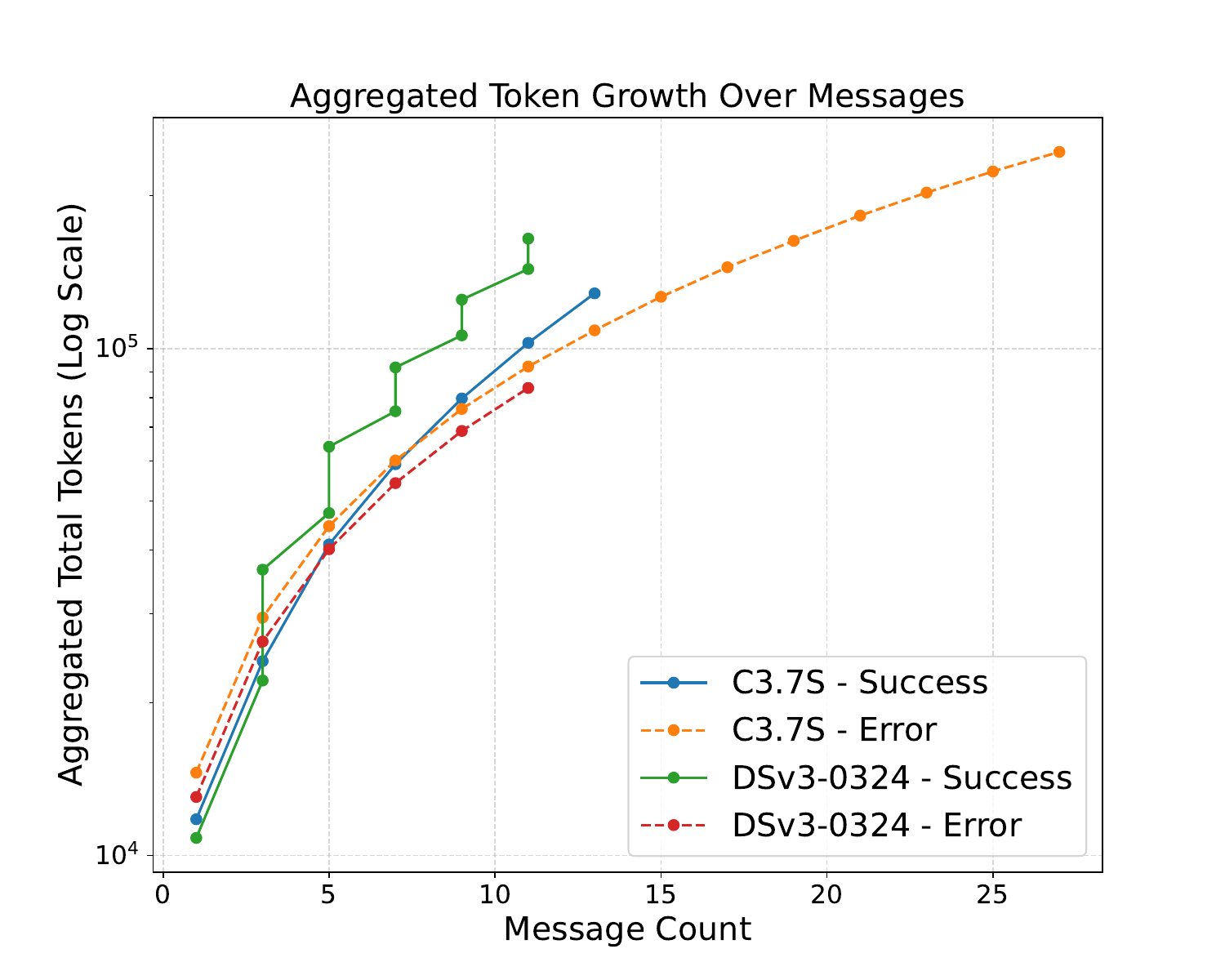}
	\vspace{-1.25 em}
	\caption{Token growth after failed tool calls.}
	\label{fig:agg_token_growth}
	\vspace{-1em}
\end{figure}

\subsection{Multi‐Tool Collaboration}
\label{subsec:multi_tool_collab}
In our benchmarks, tasks that involve multiple MCP servers are executed strictly sequentially: the LLM issues one tool call, awaits its result, and only then issues the next, even when calls are independent. This serial workflow increases both prompt size—due to repeated context serialization—and end‐to‐end latency. We therefore propose extending MCP Hosts and Clients to support parallel or batched tool invocations. By decomposing a task into independent MCP calls that can be dispatched concurrently, hosts can reduce total token exchange and cumulative latency, yielding more efficient execution.

\subsection{Prompt Token Overhead of Tool Integration}
\label{subsec:results_tokens}

Our experiments reveal that MCP‐driven tasks incur significantly higher prompt overhead than baseline chat. 
The breakdown in Figure~\ref{fig:cline_context} shows that repeated serialization of the system prompt, tool schemas, interaction history, and recent outputs drives most of this inflation.
We also find that prompt size grows roughly linearly with the number of enabled tools. To prevent context‐window exhaustion, platforms like Cursor and Cline cap concurrent tools at 40. 

Moreover, error loops, where failed tool calls trigger full re-appends of history and system prompts, can cause runaway token growth (shown in Figure~\ref{fig:agg_token_growth}), rapidly exceeding context limits and inflating costs. We therefore recommend lightweight checkpointing or loop‐detection modules to abort and restart stalled tasks.  
 
\label{subsec:results_factors}

\section{Related Work}
\label{sec:related}
Existing surveys on the Model Context Protocol (MCP) have focused primarily on its design and ecosystem rather than its performance characteristics. Singh \emph{et al.} \cite{202504.0245} present a comprehensive architectural overview of MCP, detailing its client–server model, standardized messaging, dynamic tool discovery, and potential applications across domains such as finance and healthcare. However, this work does not evaluate the token or latency overhead introduced by MCP in practice.

Complementing this, Hou \emph{et al.} \cite{hou2025modelcontextprotocolmcp} analyze the MCP landscape through a security lens, examining threats at each phase of the MCP server lifecycle, including creation, operation, and update, and proposing mitigation strategies for installer spoofing, sandbox escape, and configuration drift. While they identify critical risks and outline a roadmap for secure MCP deployment, they do not address performance trade-offs or quantify the impact on model-agent interactions.

In contrast to these studies, our work is the first to provide a systematic, measurement-based evaluation of MCP-enabled agents, quantifying prompt-to-completion token inflation, latency increases, and cost implications across real-world and controlled benchmarks.

\section{Conclusion}
\label{sec:conclusion}

In this paper, we have presented the first systematic, measurement study of the performance trade-offs inherent in MCP-enabled LLM interactions. By combining a usage trace from OpenRouter with a controlled benchmark using Cline as an instrumented MCP host, we quantified the prompt-to-completion token inflation—ranging from 2x to 30x compared to baseline chat across nine state-of-the-art LLM models. Our multi-dimensional analysis further revealed that this token overhead directly translates into higher monetary cost and increased latency, particularly in complex, multi-tool workflows.

\section{Limitations and Release Rationale}
All experiments in this study were completed on or before \textbf{May 15, 2025}, with the 90-day general-usage trace spanning \textbf{Feb 5 -- May 5, 2025}. 
Given the rapid evolution of the Model Context Protocol ecosystem, including its hosts, servers, and LLM APIs, the results presented herein should be considered a snapshot of the environment at the time of our study.
While subsequent work has advanced the state of the art, the methods and findings presented here remain relevant.
We have chosen to release this version on arXiv to establish a timestamped record of our methodology and results, thereby supporting future efforts in reproducibility and longitudinal studies.
The instrumented Cline host, benchmark scripts, and all related artifacts used in this study are available from \href{mailto:zd75@rutgers.edu}{\texttt{zd75@rutgers.edu}} upon request.

\section{Ethics}
This work does not raise any ethical issues.

%%%%%%%%%%%%%%%%%%%%%%%%%%%%%%%%%%%%%%%%%%%%%%%%%%%%%%%%%%%%%%%%%%%%%%%%%%%%
% We're in the endgame now

\bibliographystyle{ACM-Reference-Format}
\bibliography{refs}

\clearpage
\appendix

\section{Appendix: Benchmark Prompts and Task Definitions}
\label{sec:appendix_prompts}

This appendix provides extended visualizations referenced in Section~5.2. 
The visualizations include a detailed total-token breakdown and per-model radar charts that present trade-offs across four normalized metrics: task success, total tokens, latency, and monetary cost. 
(For layout reasons and to avoid large white space on the page, the total-token figure is placed immediately below this paragraph; the radar charts follow.)

The second part of this appendix lists the full set of benchmark tasks and prompts used in our MCP evaluation, provided to ensure full transparency and reproducibility of the experiments.

% ---------- Total-token figure
\begin{figure*}[!b]
  \centering
  \includegraphics[width=\textwidth]{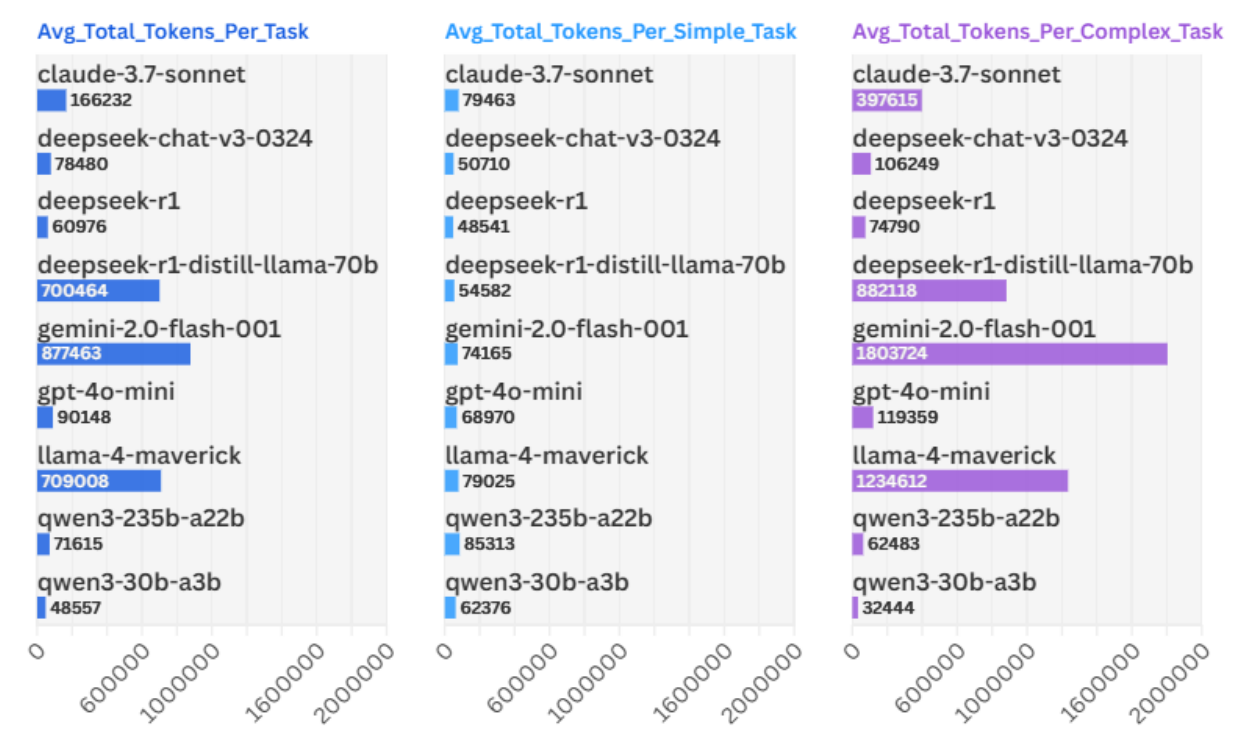}
  \caption{Average total tokens per task across evaluated models. 
  Left: all tasks; Center: simple tasks; Right: complex tasks. 
  Complex tasks amplify completion-token growth for several models, increasing both cost and latency.}
  \label{fig:total_tokens_per_task}
\end{figure*}

% ---------- Extended visualizations
\begin{figure*}[htbp]
  \centering
  \includegraphics[width=0.90\textwidth]{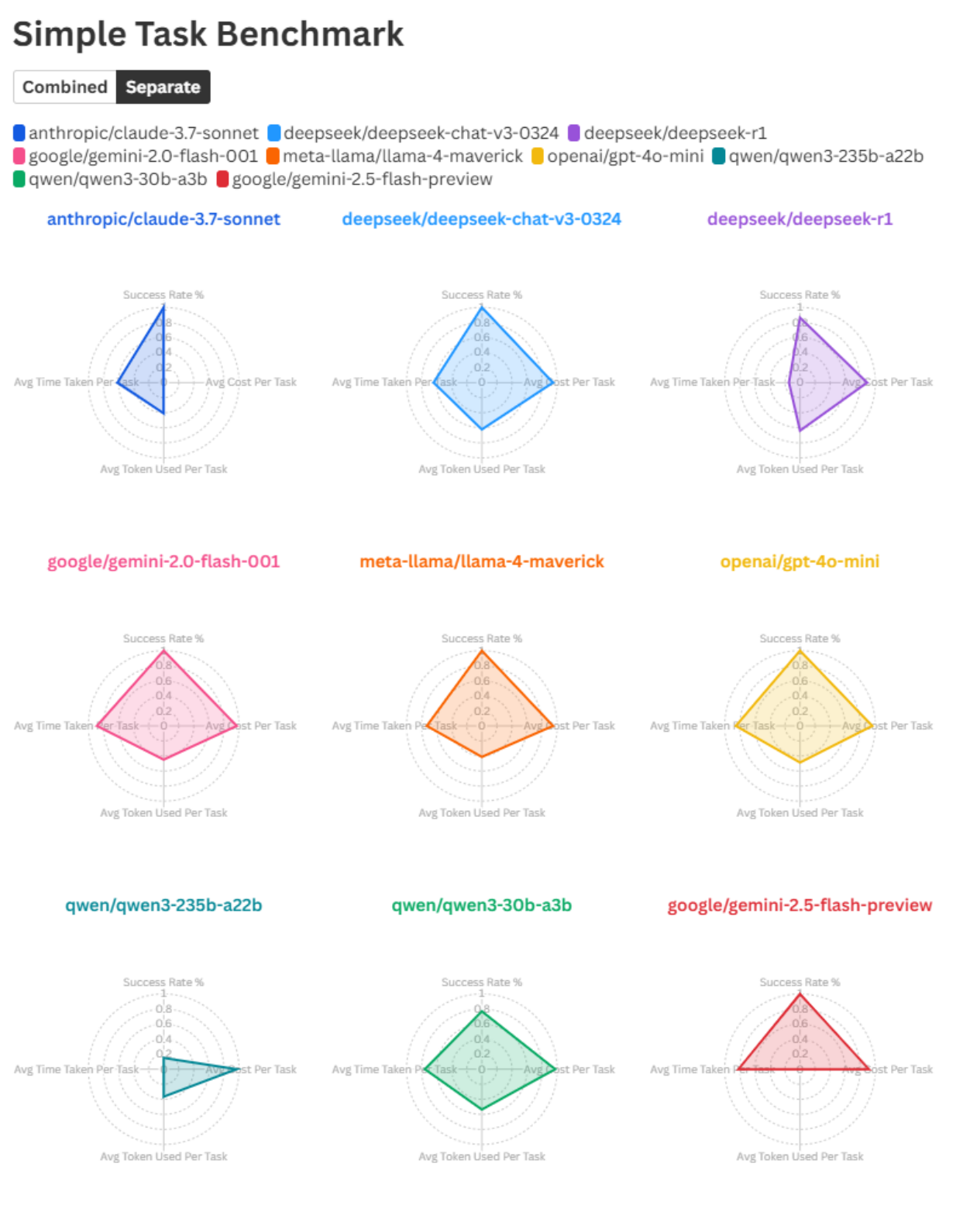}
  \caption{Per-model radar charts for the \emph{simple-task} benchmark. 
  Each radar shows normalized \textbf{Success}, \textbf{Tokens} (prompt{+}completion), \textbf{Time} (latency), and \textbf{Cost}. 
  Most models achieve high success on simple tasks, with varying token and cost footprints.}
  \label{fig:simple_radar}
\end{figure*}

\begin{figure*}[htbp]
  \centering
  \includegraphics[width=0.90\textwidth]{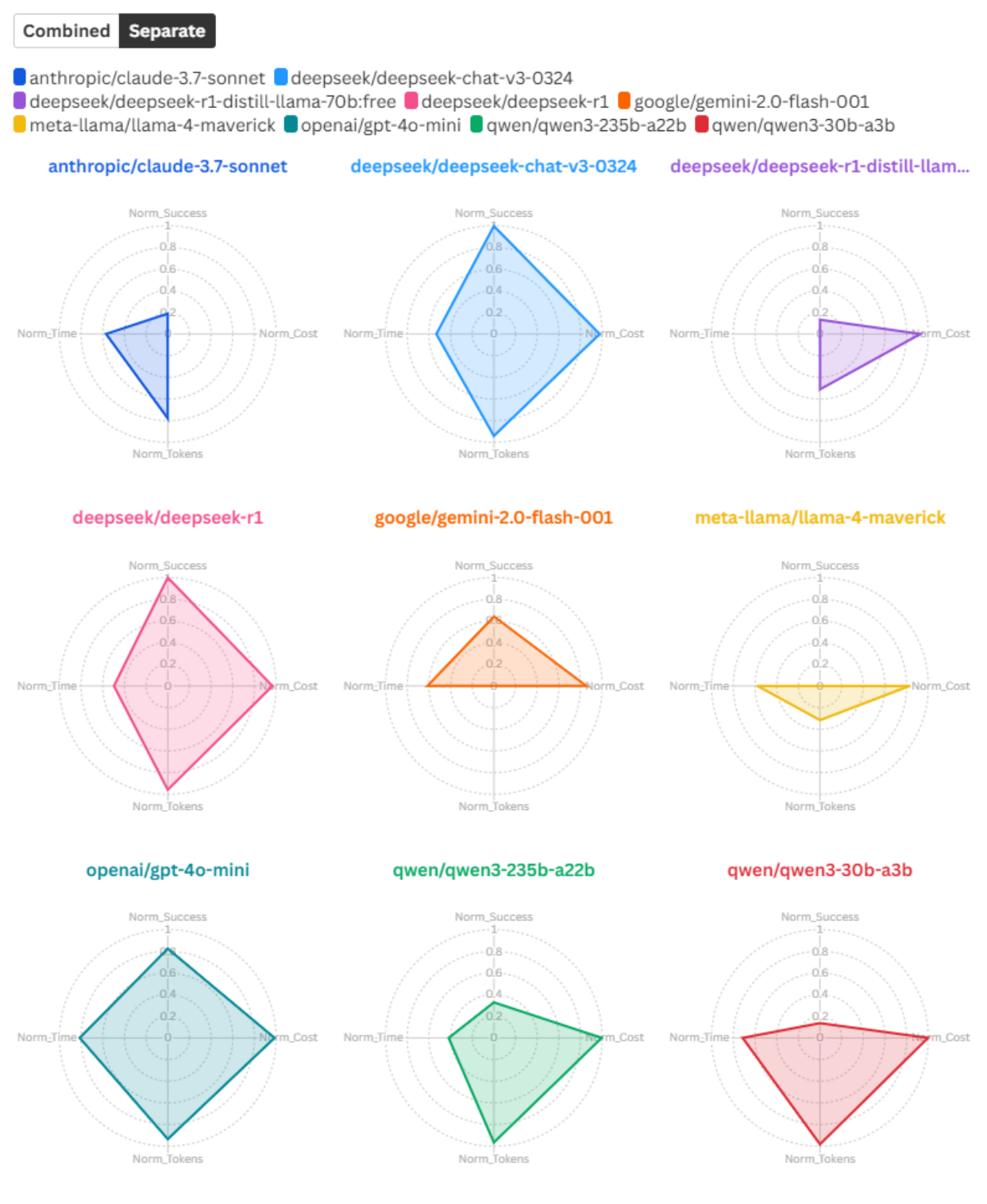}
  \caption{Per-model radar charts for the \emph{complex-task} benchmark (extension of Section~5.2). 
  Complex tasks expose larger trade-offs between effectiveness (success) and resource usage (tokens, time, cost).}
  \label{fig:complex_radar}
\end{figure*}

\begin{table*}[]
\centering
\caption{Benchmark Task Definitions and Prompts}

\label{tab:benchmark_tasks}
\begin{tabular}{@{}p{2.8cm} p{1.6cm} p{1cm} p{1.1cm} p{10cm}@{}}
\toprule
\textbf{Task Name} & \textbf{MCP Server(s)} & \textbf{Min. MCP Calls} & \textbf{Task Difficulty} & \textbf{Prompt} \\
\midrule
Get Current Time & Time MCP & 1 & Simple & What time is it?\newline \\
Time Conversion & Time MCP & 2 & Simple & Using Time MCP, first get the current date and time in the \textquotesingle{}America/New\_York\textquotesingle{} timezone. Then, convert this time to \textquotesingle{}Asia/Shanghai\textquotesingle{} using Time MCP. Finally, provide a single sentence summarizing the current time in both \textquotesingle{}America/New\_York\textquotesingle{} and \textquotesingle{}Asia/Shanghai\textquotesingle{} timezones. Ensure you achieve this without any user input.\newline \\
Where Pizza Simple & Google Map MCP & 2 & Simple & Using Google Maps MCP, search for the nearest pizza shops around \textquotesingle{}Rutgers University Busch Student Center, NJ 08854\textquotesingle{} within a 5-mile radius. From the search results, identify and rank the three closest shops. Provide their names, addresses and ratings in a clear list in the chat. Complete this task without asking me for any input.\newline \\
Where Pizza & Google Map MCP & 5 & Simple & Using Google Maps MCP, search for the nearest pizza shops around \textquotesingle{}Rutgers University Busch Student Center, NJ 08854\textquotesingle{} within a 5-mile radius. From the search results, get the place details of the three closest shops. Finally, provide their names, addresses, contact information (if available), and ratings in a clear list. Complete this task without asking me for any input.\newline \\
Sum AI News & Playwright MCP & 2+ & Complex & Summarize AI news for me related to MCP on bing news. Open the first news link, and summarize the main points in 2-3 sentences. Prioritize using Playwright MCP for interacting with the website. Try to complete this task without asking for any input from me.\newline \\
Weather 3 Days Ago & Time MCP,\newline Playwright MCP & 2+ & Complex & Using Time MCP, first get the current date and time. Then, using the \textquotesingle{}Weather Underground\textquotesingle{} website, find the temperature and weather condition for New York City exactly three days ago. Finally, provide a single sentence that includes the date (three days ago), the temperature, and the weather condition.\newline \\
Which Coffee & Playwright MCP & 2+ & Complex & Find the lowest-priced Hazelnut Ground Coffee available. Provide the name, price, and store name. Avoid user input entirely.\newline \\
Currency Conversion & Time MCP, Playwright MCP & 2+ & Complex & Using XE.com, get the current exchange rate for USD to RMB. Provide the exchange rate and timestamp.\newline \\
\bottomrule
\end{tabular}
\end{table*}

\end{document}